# Investigation of the magnetic ground state of the ordered double perovskite Sr$_2$YbRuO$_6$: a tale of two transitions

Shivani Sharma[1, 2*], D. T. Adroja[1, 3$], C. Ritter[4], D. Khalyavin[1], P. Manuel[1], Gavin B. G. Stenning[1], A. Sundaresan[2], A. D. Hillier[1], P. P. Deen[5], D. I. Khomskii[6] and S. Langridge[1],

[1]*ISIS facility, Rutherford Appleton Laboratory, Chilton Oxon, OX11 0QX*
[2]*Jawaharlal Nehru Centre for Advanced Scientific Research, Jakkur, Bangalore, India*
[3]*Highly Correlated Matter Research Group, Physics Department, University of Johannesburg, Auckland Park 2006, South Africa*
[4]*Institut Laue-Langevin, 71 Avenue des Martyrs, CS 20156, 38042, Grenoble Cedex 9, France*
[5]*European Spallation Source ERIC, 22363 Lund, Sweden*
*Niels Bohr Institute, University of Copenhagen, Denmark*
[6]*II Physikalisches Institut, Universitaet zu Koeln, Zuelpicher Str. 77, 50937 Koeln, Germany*
(Date: 18-09-2020)

**Abstract:**

Comprehensive muon spin rotation/relaxation (μSR) and neutron powder diffraction (NPD) studies supported via bulk measurements have been performed on the ordered double perovskite Sr$_2$YbRuO$_6$ to investigate the nature of the magnetic ground state. Two sharp transitions at $T_{N1}$ ~ 42 K and $T_{N2}$ ~ 36 K have been observed in the static and dynamic magnetization measurements, coinciding with the heat capacity data. In order to confirm the origin of the observed phase transitions and the magnetic ground state, microscopic evidences are presented here. An initial indication of long-range magnetic ordering comes from a sharp drop in the muon initial asymmetry and a peak in the relaxation rate near $T_{N1}$. NPD confirms that the magnetic ground state of Sr$_2$YbRuO$_6$ consists of an antiferromagnetic (AFM) structure with interpenetrating lattices of parallel Yb$^{3+}$ and Ru$^{5+}$ moments lying in the *ab*-plane and adopting a A-type AFM structure. Intriguingly, a small but remarkable change is observed in the long-range ordering parameters at $T_{N2}$ confirming the presence of a weak spin reorientation (i.e. change in spin configuration) transition of Ru and Yb moments, as well as a change in the magnetic moment evolution of the Yb$^{3+}$ spins at $T_{N2}$. The temperature dependent behaviour of the Yb$^{3+}$ and Ru$^{5+}$ moments suggests that the 4*d*-electrons of Ru$^{5+}$ play a dominating role in stabilizing the long range ordered magnetic ground state in the double perovskite Sr$_2$YbRuO$_6$ whereas only the Yb$^{3+}$ moments show an arrest at $T_{N2}$. The observed magnetic structure and the presence of a ferromagnetic interaction between Ru- and Yb- ions are explained with use of the Goodenough-Kanamori-Anderson (GKA) rules. Possible reasons for the presence of the second magnetic phase transition and of a compensation point in the magnetization data are linked to competing mechanisms of magnetic anisotropy.

## I. Introduction

Mixed ruthenates with perovskite-based crystal structures have been receiving considerable attention in recent decades[1–8], because of their interesting magnetic properties including the recent discovery of spin-triplet superconductivity in the layered ruthenate $Sr_2RuO_4$[9]. Despite the rarity of 4$d$-based magnetic materials, $SrRuO_3$ has a robust Curie temperature $T_C \sim 165$ K with saturation magnetization value of 1.4 $\mu_B$/Ru and a metallic ground state[10], while $SrRu_2O_6$ exhibits antiferromagnetic (AFM) ordering at $T_N = 563$ K and has a semiconducting ground state[11]. $Sr_2YRuO_6$, which has essentially the same crystal structure as $SrRuO_3$, but with every second Ru substituted by Y, orders in an AFM structure with an insulating ground state[1,3]. Interestingly the estimates of the ordered Ru moments is even higher than those of the parent compound, although the critical temperature is strongly reduced to 32 K ($T_{N1}$) with a second AFM transition $T_{N2} = 24$ K[1,3].

The first detailed study of the $M_2RE$RuO$_6$ ($M$ = Ca, $RE$ = Y, La, or Eu; $M$ = Sr, $RE$ = Y; $M$ = Ba, $RE$ = La or Eu) ruthenium perovskites was carried out by Greatrex $et\ al.$[12], who determined the crystal structure, measured the temperature dependence of the electrical resistivity and of the magnetic susceptibility and the $^{99}$Ru Mössbauer effect at 4.2 K. They reported that these materials crystallize in the monoclinic $P2_1/n$ space group and are magnetically ordered at 4.2 K, with $T_N$ ranging from 12 K for $Ca_2LaRuO_6$ to <80 K for $Ba_2LaRuO_6$, with hyperfine magnetic fields $B_{hf}$ at the Ru sites between 56 -60 T due to the electronic magnetic ordering[12]. In subsequent years, the AFM ordered Ru-based double perovskites, $Sr_2RE$RuO$_6$ ($RE$ = rare-earth Ho, Tb, Yb, Dy and Lu or Y etc.) were reported to exhibit two magnetic transitions and strong geometrical frustration above the magnetic ordering for some of these systems confirmed via bulk and microscopic measurements[3–6,8]. Recent neutron diffraction studies for $RE$ = Y allowed to understand and differentiate the origin of the two magnetic transitions[3] whereas for $RE$ = Dy, Ho and Tb, the difference between the two magnetic transitions could not be resolved in the neutron diffraction study within the available instrumental resolution[5,8]. In $Sr_2YRuO_6$, only half of the Ru-layers order magnetically below $T_{N1}$ while the other half (alternately) reveals short-range ordering. Furthermore, below $T_{N2}$, the system exhibits a fully ordered type-I AFM ground state[3]. The cubic double perovskite $Ba_2YRuO_6$ with space group $Fm$-$3m$ also exhibits two apparent transitions at 47 and 36 K and type I AFM ground state at low temperature[13]. Polarized neutron diffraction data revealed that this regime between 36 and 47 K is dominated by short-range spin correlations. However, the origin of $T_{N2}$ in some of these double perovskites with type-I AFM structure below $T_{N1}$ is still an open question[4,5,14,15] and the aim of the present work is to develop better understanding using the experimental data which could help to resolve this enigma for $Sr_2YbRuO_6$.

Earlier assumptions that the two magnetic transitions in $Sr_2YbRuO_6$ are due to the ordering of Yb and Ru moments at different temperatures seems unlikely due to the presence of two such transitions in the $Sr_2YRuO_6$ where only one magnetic cation (i.e. Ru) is present[3,4]. Further intriguing facts regarding the magnetic ground state of the Ru- based double perovskites are the similar ordered moment values (~2 $\mu_B$) found for the $Ru^{5+}$ ion irrespective of the nature of the $RE$ (rare earth) atom and the small value of the ordered moments of the magnetic $RE$ ions[3,5,8]. All these results motivate further exploration of the other members of this family in order to understand the origin of the two magnetic transitions, the role of the Ru-atom in the magnetic ordering, and the participation of rare-earth atom in determining the magnetic ground state.

Sr$_2$YbRuO$_6$ is a magnetic insulator with a double-perovskite structure, which undergoes a long-range magnetic ordering transition below $T_{N1}$ (42 K), in addition to the conspicuous occurrence of the second transition at $T_{N2}$ =36 K and a weak anomaly at $T^*$ = 10 K[4,15]. Sr$_2$YbRuO$_6$ also displays a temperature induced magnetization reversal almost coinciding with $T_{N2}$ due to an underlying magnetic compensation phenomenon[16]. The observed magnetic entropy $S_{mag}$ = 5.7 J mol$^{-1}$ K$^{-1}$ at 60 K is smaller than the expected value for ordered Ru$^{5+}$ moments with a ground state of $J$ = 3/2 ($S_{mag}$ = 11.52 J mol$^{-1}$ K$^{-1}$)[4]. This was tentatively linked to the presence of frustration above the magnetic transition. The same group has also reported the exchange bias effect in Sr$_2$YbRuO$_6$ below the compensation temperature[16]. The compensation temperature was referred to as the temperature where the measured magnetization becomes zero[4] and a cross-over of zero field cooled and field cooled magnetization occurs. However, in the same report, it was suggested that two magnetic anomalies near $T_{N1}$ and $T_{N2}$ could be due to the magnetic ordering of Ru$^{5+}$ (4d$^3$) and Yb$^{3+}$ (4f$^{13}$) moments, respectively. Later, Doi et al.[15] reported a type-I AFM structure below $T_{N1}$ confirmed via neutron powder diffraction (NPD) study performed at 10 K. However, due to the lack of systematic temperature dependent NPD data, no information is available regarding the thermal evolution of the magnetic structure at $T_{N2}$[14,15]. Here, we present a detailed NPD and μSR study, which, supported by exhaustive magnetisation and heat capacity data, confirms that both the Ru$^{5+}$ and Yb$^{3+}$ moments order at $T_{N1}$ and that a weak spin reorientation takes place at $T_{N2}$. We use this term "spin reorientation" in the sense of "change in the relative spin configuration". No change or anomaly has been found near $T^*$~ 10 K in the NPD data.

**II. Experimental details**

The polycrystalline sample of Sr$_2$YbRuO$_6$ was prepared by the standard solid-state reaction using the same protocol as mentioned elsewhere[4]. Phase purity was confirmed by X-ray diffraction (XRD) using a Rigaku Smartlab X-ray diffractometer equipped with a Ge two bounce monochromator enabling Cu-Kα radiation. The *dc* magnetization measurements have been performed on a Quantum Design's SQUID magnetometer. Temperature dependent heat capacity, using a relaxation technique, and *ac*-susceptibility were measured using PPMS by Quantum Design. To investigate the magnetic structure/ground state, temperature dependent NPD measurements were carried out using the time-of-flight diffractometer WISH at the ISIS Facility, UK[17]. The FullProf_Suite has been used to analyse the XRD and NPD data[18]. The MuSR spectrometer in longitudinal geometry at the ISIS Pulsed Neutron and Muon Source, UK has been employed to carry out zero-field (ZF) μSR experiments. The powder sample was mounted onto a silver plate (99.999% purity) using GE-varnish and was covered with thin silver foil. The μSR measurements were carried out using He$^4$ cryostat between 2 and 300 K.

**III. Results and discussion**

**(a) X-ray diffraction**

The room temperature XRD pattern of Sr$_2$YbRuO$_6$ has been Rietveld refined using monoclinic symmetry (space group $P2_1/n$) with an ordered arrangement of Yb$^{3+}$ and Ru$^{5+}$ atoms at the B-site. The result is shown in Fig. 1(a) and is in good agreement with the existing literature[4,14]. No extra peaks were evident in the XRD pattern while a very minute impurity phase of Yb$_2$O$_3$ was evident in NPD pattern. One can easily miss out this minute impurity with a lab source based XRD machine while the high intensity available of the neutron beam on the WISH instrument, this minute phase can be easily seen. The results of NPD will be discussed in later sections. It is imperative to mention that the magnetic ordering of the impurity phase $Yb_2O_3$ cannot be responsible for the appearance of *T\** as its transition temperature is much lower at about 2.25 K[19]. The crystal structure and its details of $Sr_2YbRuO_6$ are presented in Fig. 1(b). Bond lengths and bond angles governing the different magnetic interaction pathways are shown in the enlarged views of two dashed box regions: Fig. 1(c) and (d).

**(b) ac- and dc-Magnetization**

Figure 2(a) displays the zero-field cooled (ZFC) and field cooled (FC) *dc* magnetization ($\chi_{dc}$) behaviour of $Sr_2YbRuO_6$ measured in different fields namely, at 50 Oe, 100 Oe and 10 kOe as a function of temperature. The bifurcation of the ZFC and FC magnerization only starts below a certain critical temperature, following by a crossover between the ZFC and FC curve. For low applied fields (50 and 100 Oe), the FC magnetization becomes negative by cooling the sample below the crossover point, whereas for sufficiently high fields (10 kOe), the FC curve stays always positive. Noticeably, the ZFC magnetization decreases below 42 K showing a plateau for a small temperature region down to 36 K. Below 36 K the ZFC, magnetization increases with decreasing temperature, irrespective of the applied field value. Here we denote these anomalies by $T_{N1}$ (42 K) and $T_{N2}$ (36 K), respectively. The justification and microscopic evidence to denote them as AFM ordering temperatures ($T_N$'s) comes from the NPD results which are discussed later. Another intriguing feature is the presence of a weak anomaly near 10 K. A similar anomaly below 15 K was previously mentioned to exist by Singh *et al.* [4]. We denote this anomaly by *T\**, as we do not have any existing information about its origin. Both the ZFC and FC curves exhibit a small kink near *T\**.

To investigate further the nature of these magnetic anomalies, isothermal magnetization has been measured at selected temperatures. Figure 2(b) represents the magnetic isotherms measured at *T* = 5, 30, 37, 50 and 300 K. Noticeably, a weak hysteresis starts to develop below 37 K and becomes quite prominent for the 5 and 30 K curves. It suggests the contribution of a minor ferromagnetic (FM) component to the dominant AFM ground state. The 300 K and 50 K curves exhibit a linear behaviour, as expected for a paramagnetic state.

To understand the dynamic response of these anomalies, the *ac* susceptibility ($\chi_{ac}$) of $Sr_2YbRuO_6$ has been measured. Figure 3 represents the real ($\chi'$) part of $\chi_{ac}$ as a function of temperature measured at different frequencies. Two clear anomalies are visible in the $\chi'$ behavior near $T_{N1}$ and $T_{N2}$. The frequency independent behaviour of the first anomaly at $T_{N1}$ indicates the onset of long-range ordering below $T_{N1}$ as shown in the enlarged view as inset (i) of Fig.3. A weak frequency dispersion can be seen below $T_{N2}$, which indicates the change in magnetic interactions at this point, shown in inset (ii) of Fig.3. A similar kind of frequency dispersion at $T_N$ has also been observed for other systems showing long-range ordered state, for example $Sr_3NiIrO_6$ and $Sr_2DyRuO_6$ (near $T_{N2}$)[8,20,21]. A very weak, indirect but apparent signature of a third anomaly near *T\** can be seen in the $\chi_{ac}$ behaviour at T = 10 K. The frequency dispersion decreases below *T\** and $\chi'$ increases sharply. The direct signatures of $T_{N1}$ and $T_{N2}$ have been also found in the $\chi''$ behavior but due to the weak signal, it is difficult to find any signature of *T\** in $\chi''$ behavior (data is not shown here).

**(c) Heat capacity**

The heat capacity of $Sr_2YbRuO_6$ measured in 0 and 2 T applied field is presented in Fig. 4. Two clear peaks are visible near 42 and 36 K, coinciding with the magnetic anomalies at $T_{N1}$ and $T_{N2}$, respectively, which confirms the long-range ordering at these transitions. However, no feature or anomaly has been observed near $T^*$. Also, there is no appreciable change in the heat capacity behaviour measured with 0 and 2 T applied field (Fig. 4). Therefore, the static and dynamic magnetization and heat capacity measurements confirm the presence of two long-range transitions at $T_{N1}$ and $T_{N2}$.

**(d) Muon spin rotation and relaxation**

In order to understand the microscopic origin and local magnetic response of different phase transitions as observed through the bulk techniques, the zero field (ZF) µSR spectra of $Sr_2YbRuO_6$ have been recorded at various temperatures between 2 and 90 K as shown in Fig. 5. The spectra at 90 and 50 K exhibit weak relaxation and have the same initial asymmetry. However, below 45 K, the relaxation rate increases faster and the initial asymmetry decreases with decreasing temperature. This is a typical behaviour observed near a long-range magnetic ordering transition. The ZF µSR data is fitted using an exponential function with a constant background.

$$G_z(t) = A_0 \exp(-\lambda t) + A_{bg} \qquad (1)$$

Here $A_0$ is the muon initial asymmetry, $\lambda$ the muon relaxation rate, $A_{bg}$ is the constant background arising from muons stopping on the sample holder. The value of $A_{bg} = 0.02$ was estimated from the fitting of the 90 K data and then kept fixed for fitting the data at other temperatures. The fitting parameters, relaxation rate ($\lambda$) and initial asymmetry ($A_0$) are plotted in Fig. 6. For temperatures down to 50 K, the initial asymmetry is almost temperature independent, which can be attributed to fluctuations of the paramagnetic moments of the $Yb^{3+}$ and $Ru^{5+}$ ions. $\lambda(T)$ increases below 50 K and exhibits a sharp maximum near 42 K ($T_{N1}$). At $T_{N1}$, the initial asymmetry drops down by more than 2/3 of the initial value, which indicates that the magnetic ordering is bulk in nature. In a polycrystalline sample, below the magnetic ordering temperature, muons see three components (one longitudinal and two transverse) of the internal field at muon stopping sites. For a bulk magnetic ordering with larger magnetic moments one expects a 2/3 loss of initial asymmetry (the 2/3 transverse component gives oscillations and the 1/3 longitudinal component gives a relaxation) as the transverse component can be seen only very close to the zero-time limit for larger internal fields at muon stopping sites. In the present case, the asymmetry loss is slightly larger than 2/3, which could be due to a fast relaxing component below $T_{N1}$ at smaller time, which cannot be estimated due to the muon pulse width (70 ns FWHM) at ISIS. For $T < T_{N1}$, the further loss in initial asymmetry is very small while the relaxation rate $\lambda(T)$, after peaking at $T_{N1}$, continues to decrease down to lowest temperatures. As expected $A_0$ does not reveal any sign of a second/third transition as the system is in a complete long-range magnetic order state below $T_{N1}$ and hence cannot lose further asymmetry. It is interesting to notice that the observed maxima/peak in $\lambda(T)$ near $T_{N1}$ agrees with the susceptibility and heat capacity data. However, the continuous change of $\lambda(T)$ across $T_{N2}$ and $T^*$ indicates a small change in the magnetic structure specifically at $T_{N2}$. Similar kind of responses have been recently observed for various other perovskites[7,22–25] and have been helpful in exploring the magnetic ground states, including the microscopic co-existence of magnetic ordered and non-magnetic phases in Ba$_2$PrRu$_{0.9}$Ir$_{0.1}$O$_6$ using μSR[26].

**(e) Neutron diffraction**

To investigate the magnetic ground state and the possible changes in the magnetic structure across the different transitions, NPD data have been collected on the WISH time-of-flight diffractometer at several temperatures between 100 to 2 K with close data points between 45 and 2 K. The emergence of new peaks along with the enhancement in the intensity of some nuclear peaks is clearly observed below $T_{N1}$. Fig. 7 represents in a 3D plot the thermodiffractogram of Sr$_2$YbRuO$_6$ for $T < 45$ K and interplanar spacing $d > 3.5$ Å. All the magnetic reflections can be indexed with a propagation vector $k = (0,0,0)$. The occurrence of the (010) reflection indicates that the magnetic moments should have components perpendicular to the $b$-direction. No additional magnetic Bragg peaks appear below $T_{N2}$ or below $T^*$. The black and red arrows in Fig. 7 point to the temperatures corresponding to $T_{N1}$ and $T_{N2}$. A non-monotonic change of the intensities of the magnetic Bragg peaks is visible at $T_{N2}$ and suggests a change in the magnetic structure across $T_{N2}$. However, a detailed Rietveld refinement is needed to confirm and describe these changes of the magnetic structure at $T_{N2}$; this will be discussed below. A qualitative representation is given in Fig. 8 where the thermal evolution of the integrated intensity of various magnetic reflections is plotted against temperature. All reflections exhibit a first rise below $T_{N1}$ concomitant with the onset of long-range ordering. Below $T_{N2}$, they exhibit a more or less pronounced accelerated enhancement in the diffracted intensity with decreasing temperature. Since, all the observed magnetic Bragg peaks can be fitted with the type-I AFM structure (which is discussed below in detail), the observation of two different temperature regions in the thermal behavior of the magnetic reflections can explain the existence of two peaks in the magnetization and the heat capacity behavior. The red lines in Fig. 8 are guides to the eye for the expected temperature variation of moment components arising below $T_{N1}$ and $T_{N2}$. The temperature evolution of (010) and (100/001) peaks in Fig. 8 clearly supports the presence of two magnetic transitions. It is to be noted that no further deviation or anomalous change in the long-range order parameter (integrated intensity) has been observed at $T^*$ in Sr$_2$YbRuO$_6$. A similar two-step behavior (at $T_{N1}$=31.9 K and at $T_{N2}$=24 K) in the intensity of the magnetic Bragg peaks was also observed for Sr$_2$YRuO$_6$[3] and has been interpreted as corresponding to a 2D magnetic transition (where only half the Ru planes ordered magnetically) at $T_{N1}$ followed by a 3D magnetic transition (all Ru atoms order magnetically) at $T_{N2}$. Unfortunately no details on the space group used in the analysis of the magnetic structure of Sr$_2$YRuO$_6$ or of the Wyckoff positions of the Ru atoms used during the refinement of the neutron diffraction data were given. The idea of having only half of the Ru layers magnetically ordered below $T_{N1}$ while all Ru-layers become magnetic ordered below $T_{N2}$ demands the existence of two different crystallographic sites for the Ru atoms in the crystal structure of Sr$_2$YRuO$_6$. This is not the case in the normally used space group $P2_1/n$ where only one crystallographic Ru site exists.

To investigate the corresponding changes in the magnetic structure of Sr$_2$YbRuO$_6$, Rietveld refinements were done using the total diffracted intensities and the temperature dependent difference data sets where the nuclear contribution using the 45 K data set had been subtracted. The difference data sets are more sensitive to small changes of the magnetic structure expected to happen at $T_{N2}$. All the five banks of data have been refined simultaneously to get the final parameters. Fig. 9 represents the Rietveld refined plot of the 100 K (a) and 2 K (b) data from the bank 2 of WISH instrument. The 100 K data were fitted with a nuclear phase having the monoclinic space group $P2_1/n$. A very minute (~1.5 %) impurity of $Yb_2O_3$, which orders at 2.25 K[19], was found as well in the NPD pattern. The 2 K data are fitted using a two-phase (nuclear + magnetic) model. The inset in Fig. 9(b) shows the Rietveld refined plot of the difference data at (2 K − 45 K) fitted only with the magnetic phase using a fixed scale factor determined from the refinement of the purely nuclear data at T = 45 K. The refined lattice parameters at 2 K are $a$ = 5.7305(2) Å, $b$ = 8.1021(3) Å, $c$ = 5.7360(2) Å and γ = 90.20(2)°. It should be noted here that we have used the $P112_1/n$ setting instead of the standard $P12_1/n1$ ($a$ = 5.7314(2) Å , $b$ = 5.7367(1) Å , $c$ = 8.1029(3) Å , β= 90.182(1) ° at 100 K) used in the previous work Doi et al.[15], because the former gives an advantage to adopt the polar coordinates during the refinement procedure. The empirically determined magnetic form factor of $Ru^{5+}$ has been used for the refinement[27]. Magnetic symmetry analysis was performed using the space group $P2_1/n$ with k = (0, 0, 0) using the program BASIREPS[28] which generates two possible irreducible representations (IR1 and IR2), each containing three basis vectors. IR1 has ferro- (F) coupling along the $c$-direction and antiferro- (AF) coupling in the $a$ and $b$-directions while on the contrary, the IR2 has AF coupling in the $c$-direction and F-coupling in the $a$ and $b$-directions. The best fit of the data can be achieved with a single IR1, having AF-coupling along the $a$ and $b$-direction. A collinear model, having parallel $Yb^{3+}$ and $Ru^{5+}$ moments, has been used to refine the data for the magnetic structure determination. Any attempt to avoid this constraint leads to instabilities and divergence of the refinements. The final magnetic structure presented in Fig. 10 consists of an interpenetrating lattice of canted moments of $Yb^{3+}$ and $Ru^{5+}$ ions where FM sheets extending within the $a$-$c$-plane are coupled antiferromagnetically along the $b$-direction. The spins are pointing along the long $b$-axis with an angle of ~ 45 - 51° (temperature dependent) with respect to the $a$-axis. Fig. 10(b) explains the different angles used to describe the magnetic structure. For comparison, we plotted in Fig. S1 of the Supplementary Materials, the magnetic structure of $Sr_2YbRuO_6$ in the two different settings, $P112_1/n$ and $P12_1/n$ Ref. [29]. Due to the pseudosymmetry present in the sample an equally good fit of the data can be obtained by refining the magnetic structure with AFM $b$ and $c$ components. This magnetic structure, however, would not be compatible with magnetic symmetry analysis. Doi et al.[15] have reported similar magnetic structure with Ru and Yb moments at 23° relative to the c-axis at 10 K. Due to the pseudosymmetry present and the absence of magnetic symmetry analysis they were not able to specify whether the canting angle is relative to their $a$- or $b$-axis. $Sr_2TbRuO_6$[5] and $Sr_2YRuO_6$[30] are the only other members of this family of double perovskite for which a spin canting (20°, respectively 10.5° from the long axis) is known[5]. The direction of the magnetic moments of $Yb^{3+}$ and $Ru^{5+}$ are different in the present system from those of the Ho, Tb and Dy based double ruthenates[5,8,31]. The coupling between the $Yb^{3+}$ and the $Ru^{5+}$ moments is FM whereas an AFM coupling was observed between the rare earth $RE^{3+}$ and $Ru^{5+}$ for RE = Ho, Tb, Dy and Tm. While the spins are canted from the long axis in $Sr_2YbRuO_6$, in $Sr_2DyRuO_6$ both the Dy and Ru spins are at 90° to the long axis (i.e. in the plane)[8], while in $Sr_2TmRuO_6$ both the Tm and Ru spins are strictly pointing along the long axis[15]. The values of the $Yb^{3+}$ and the $Ru^{5+}$ moments at 2 K, obtained in this work, are $\mu_{Yb3+}$ = 0.54(1) $\mu_B$ and $\mu_{Ru5+}$ = 2.10(1) $\mu_B$. The strong reduction of $\mu_{Yb3+}$ compared to the expected value of ~ 4.5 $\mu_B$ matches with similar discrepancies observed for rare-earths' moment for the other members of Ru-based perovskites family, like $Sr_2DyRuO_6$, $Sr_2HoRuO_6$ and $Sr_2TbRuO_6$ etc.[5,8,15,31]. A reduced RE moment is frequently assigned to the effect of the crystal field on the rare-earth cation and/or due to the non-developed RE-RE direct magnetic exchange.

Magnetic symmetry analysis allows a FM component in IR1 on the $Yb^{3+}$ and the $Ru^{5+}$moments along the $c$-direction. The expected FM contribution to the magnetic Bragg peaks comes, however, on top of the nuclear peaks. The intensity of the nuclear peaks is determined by the atom coordinates, the B-factor (thermal factor) etc., which can all change slightly with temperature. Our attempt to determine the FM component gave very large errors and the results were not reliable. This is not surprising as the FM component of the moments expected from the magnetization isotherm measurements at 5 K given in Fig. 2b is very small ~0.01 $\mu_B$. Table I contains the information on the bond lengths and bond angles variation during these transitions. No sudden change or variation in the bond lengths/bond angles was noticed at $T_{N2}$ or $T^*$.

To discern the changes of the magnetic structure at $T_{N2}$, the temperature variation of the magnetic moments has been determined by Rietveld refinement of the temperature dependent difference data sets. The refined $Ru^{5+}$ and $Yb^{3+}$ moments are plotted in Fig. 11 as a function of temperature alongside with the angle $R_\Phi$, which describes the canting of the moments with respect to the *x*-axis (a-axis), and the normalized moments. The value of $R_\theta$ was kept constant and equal to 90° (see Fig.10b for the definition) reflecting the non-existence of a FM component along the *c*-direction. There are small but clear anomalies in the temperature dependence of the moments (more pronounced for $Yb^{3+}$) and in the $R_\Phi$ value at $T_{N2}$. The $Yb^{3+}$ moments show a sharp increase (similar to the $Ru^{5+}$ moments) below $T_{N1}$, but show an arrest in the slope near $T_{N2}$, before it is increasing again more strongly (almost linearly with temperature) and saturating near 10 K. The angle $R_\Phi$ which is determined by the relative sizes of the two AFM components along the *a*- and the *b*-directions also shows a continuous increase down to $T_{N2}$, below which it slightly decreases before saturating to ~50°. Also, from Fig. 11(d), it appears that below $T_{N2}$ the $Ru^{5+}$ moments attain the saturation value with a much faster rate compared to $Yb^{3+}$. Noticeably, the rate of increase of the $Ru^{5+}$ and $Yb^{3+}$ moments is different only below $T_{N2}$ while both the moments increase with the same rate between $T_{N1}$ and $T_{N2}$. Resuming the analysis of the temperature dependent refinement of the difference data sets one can say that a broad but clear peak in $R_\Phi$ along with a small plateau in the size of the $Yb^{3+}$ moments appear near $T_{N2}$. The change in the temperature variation of the normalized moments further confirms the change of the magnetic interactions leading to the magnetic structure at $T_{N2}$ in $Sr_2YbRuO_6$. In this context, it has to be noted that we have not seen any sign of elastic diffuse scattering between $T_{N1}$ and $T_{N2}$ in our WISH diffraction data of $Sr_2YbRuO_6$. This is different from the behavior observed in $Sr_2YRuO_6$[3] and in the cubic $Ba_2YRuO_6$[13] where the presence of short-range spin correlations has been observed between $T_{N1}$ and $T_{N2}$ and connected to a 2D ordering and the absence of true long range magnetic order. Differences between the RE=Y and Yb compounds could be linked to the different levels of magnetic frustration present in the $Sr_2RERuO_6$ (RE=rare earths) compounds. Using the frustration index $f = |\theta_{CW}|/T_N$[32] to quantify the frustration, values of *f* ranging from 0.5-0.7 for Gd to Er, increasing to 1.3 for Tm, 5.35 for Yb, 9.1-11.2 for Y and 11.7 for Lu can be found (Table II). High values of *f* have also been observed in $Ba_2RERuO_6$, 17 for RE=Y and 18 for RE=Lu (see Table II). This gives some indication that the magnetic frustration in $Sr_2YbRuO_6$ ($Ba_2YbRuO_6$) is reduced compared to $Sr_2YRuO_6$ ($Ba_2YRuO_6$) and could explain why the ordering at $T_{N1}$ is 3D.

The size of the $Ru^{5+}$ moment determined for $Sr_2YbRuO_6$ is very similar to the reported values for other members of this double perovskite family and points to the fact that in these systems the Ru-O-O-Ru interactions are the strongest magnetic interactions, which control the Ru ordering[3,5,8,15,32]. The very low value of $T_N = 2.3$ K for the rare-earth oxide $Yb_2O_3$ indicates that Yb-O-Yb interactions are in general very weak[19]. For the well-ordered double perovskite $Sr_2YbRuO_6$, even weaker super-super exchange Yb-O-O-Yb interactions will be present. The absence of magnetic order down to 2 K in $Sr_2YbMO_6$ (M=Nb, Ta and Sb)[33] indicates as well that Yb-Yb interactions are weak in these double perovskites. These interactions cannot be responsible for the $Yb^{3+}$ ordering at $T_{N1}$. Therefore, it appears that the Ru-O-Yb interactions have an important role in governing the magnetic ordering of the rare-earth cation $Yb^{3+}$. Noticeably, the $Yb^{3+}$ moment exhibits deviation from the mean field type behavior as a function of temperature while the $Ru^{5+}$ moment follows the mean field type behaviour down to 2 K. This indicates that in the rare-earth and ruthenium-based perovskites, the primary magnetic ordering below $T_{N1}$ is induced by the order of the 4d-electrons of $Ru^{5+}$ rather than by that of the rare-earth cation, as is also verified for $Sr_2RERuO_6$ (RE = Ho, Tb and Dy)[5,8].

### IV. Theoretical discussion

To explain the properties of $Sr_2YbRuO_6$ it is necessary to discuss the microscopic contributions determining the magnetic properties. As mentioned above, the most important one is the Ru-Ru exchange interaction. For the $t_{2g}^3$ occupation of $Ru^{5+}$ ions it is relatively straightforward to understand: there are no orbital degrees of freedom and the exchange is the same for nearest neighbours in all directions. Simple arguments, illustrated in Fig.12 (a), demonstrate that the Ru-O-O-Ru nearest neighbour (NN) exchange is AFM, in accordance with the Goodenough-Kanamori-Anderson (GKA) rules[34–36], see e.g. the discussion in [37]. Because of the $t_{2g}^3$ occupation, the AFM exchange would be the same for NN in xz and in yz planes. With AF interaction to 12 nearest neighbour Ru's one stabilises the type I magnetic structure (FM planes stacked antiferromagnetically): one has in this case 8 NN AFM pairs and only 4 NN FM ones.

Similar arguments also explain the exchange between Ru and Yb; as indicated above the direct Yb-Yb exchange is definitely much smaller and can play a role only at very low temperatures. The ground state of $Yb^{3+}$ ($4f^{13}$) in this case is the $\Gamma_6$ doublet[31], the shape of its wave function is sketched in Fig.12 (b) from ref. [38]. One sees that the occupied $t_{2g}$ orbitals of Ru are orthogonal to the $\Gamma_6$ states of Yb, i.e. the only exchange processes could be due to the virtual hopping from occupied to empty states, which, according to GKA rules, gives FM Yb-Ru coupling. This naturally explains why Yb spins are ordered parallel to the spins of ferromagnetically ordered Ru planes. Combining the Ru-O-O-Ru exchange and the Ru-Yb coupling one gets indeed exactly the magnetic structure observed in $Sr_2YbRuO_6$: type I ordering of both Ru and Yb sublattices being parallel, i.e. ferromagnetically coupled.

One of the interesting and important questions is the nature of two magnetic transitions in many ruthenium double perovskites, including $Sr_2YbRuO_6$. As one can see from Table II, more than half (5 of 9) of the known $Sr_2RERuO_6$ systems have a double transition. Sometimes it is seen in the magnetic data, sometimes in the specific heat, sometimes in both. Apparently, there is no (strong) change of the lattice at these transitions, i.e. they are of predominantly magnetic origin (although some effect on the lattice cannot be excluded, for example due to magnetostriction). Interestingly, these two transitions are seen both for magnetic RE (Dy, Ho, Yb) and for nonmagnetic ones (Lu, Y). From this we can conclude that it is predominantly the Ru subsystem which is responsible for the two transitions.

Based on experimental data there are two factors invoked, which could be responsible for the existence of two transitions. As already mentioned above, one explanation was put forward by E. Granado et al., in ref. [3]: in this paper based on neutron scattering it was concluded that in $Sr_2YRuO_6$ there appears two-dimensional ordering at $T_{N1}$, which becomes three-dimensional below $T_{N2}$. The indication for this behaviour was the presence of strong diffuse neutron scattering between $T_{N1}$ and $T_{N2}$, which the authors attributed to the absence of full 3D ordering in this temperature range, i.e. between $T_{N1}$ and $T_{N2}$. However, we observe no such elastic diffuse scattering and as such this explanation does not apply to our system.

The other effect noticed in $Sr_2YRuO_6$ by Singh et al.[2], and by Kayser et al.[30], which is also seen in our data, see Fig.11c, is the slight change of the spin direction at $T_{N2}$. This could be another reason for the second transition: it could be predominantly a spin reorientation. We have to point out here that our use of the term "spin reorientation" has to be understood as a non-monotonic change of the spin structure within the same irreducible representation. There is neither a change of the crystallographic nor of the magnetic symmetry connected to $T_{N2}$. The data of Fig. 11(c) show this change of the spin orientation in approaching $T_{N2}$. The behaviour of the magnetization in the ZFC and especially in the FC regime shown in Fig. 2a, with spin compensation, also corroborates this explanation. As such behaviour is seen in both Y and Yb systems, it is hardly connected to the direct influence of the rare earths (although the details may depend on those). Most probably, it is related to the specific characteristics of the Ru ions, namely to its single-site anisotropy and to the Dzyaloshinskii-Moriya (DM) interaction existing in the crystal structure of $Sr_2YbRuO_6$. In this sense the situation here strongly resembles that found in $YVO_3$[39,40] in which a compensation point was also seen in a system with only one type of magnetic ions (whereas the most common reason for compensation points is the interplay of two magnetic sublattices with different ions, having different magnetic moments). Such behaviour in $YVO_3$ was explained in[39,40] as an interplay of two mechanisms of magnetic anisotropy: single-site anisotropy $KS_z^2$ and DM interaction. Both mechanisms can create magnetization, which, however, can point in opposite directions, with different mechanisms dominating in different temperature intervals, which can lead to spin compensation at some temperature. We suppose that the same mechanism may operate also in $Sr_2YbRuO_6$ and could lead to both spin compensation and the appearance of the spin reorientation transition. This mechanism relies on a "fine tuning" of two mechanisms determining the spin direction, and the resulting behaviour may depend on the details of the system. As the magnetic anisotropy of a magnetic RE will probably contribute as well to the total balance, this can explain why the double transitions are seen in some materials, for some RE ions, but not in the others. We have used a point charge model of the crystal field to calculate the single ion anisotropy of the Yb ion in $Sr_2YbRuO_6$, which shows that the easy axis of the magnetization is the b-axis (i.e. along the long axis).

Thus, on the basis of the results of our experiments, and also analysing the tendencies in this whole class of materials, we suggest that the main mechanism governing the second magnetic transition in these systems is connected to details of the magnetic anisotropy of the Ru- and the RE-system and their temperature dependence. But the real proof of this picture may require additional studies on single crystals.

## V. Conclusions

We have investigated the ordered double perovskite $Sr_2YbRuO_6$ using various experimental techniques to understand the origin of two magnetic transitions. The bulk magnetization measurements of $Sr_2YbRuO_6$ reveal the presence of two clear magnetic transitions as a function of temperature, namely at $T_{N1}$ = 42 K, at $T_{N2}$ = 36 K and a very weak anomaly at $T^*$ = 10 K. The heat capacity measurements reveal a clear signature of $T_{N1}$ and $T_{N2}$ indicating the long-range ordering whereas no anomaly can be detected at $T^*$. Our detailed μSR and NPD results provide a concrete evidence of long-range magnetic ordering of both sublattices ($Ru^{5+}$ and $Yb^{3+}$) below $T_{N1}$ and a clear change in the long-range magnetic ordering parameters at $T_{N2}$. The magnetic ordering is primarily controlled by the $Ru^{5+}$ moments, but a change in the spin structure at $T_{N2}$ is confirmed based on the temperature variation of $Yb^{3+}$ and $Ru^{5+}$ moments and of the angle $R_\phi$ describing the moment direction. All the observed magnetic Bragg peaks can be indexed with a single propagation vector $k = (0, 0, 0)$ and the magnetic structure consists of interpenetrating sublattices of $Yb^{3+}$ and $Ru^{5+}$ spins having confined moments in the *ab*-plane. The resultant magnetic structure is composed of parallel spins of $Yb^{3+}$ and $Ru^{5+}$ having an angle of $R_\phi \sim 45\text{-}51°$ with respect to the *a*-direction.

Based on our and related data on similar systems, we propose that the second magnetic transition and the presence of a compensation point in the magnetization observed in many materials of this class may be related to details of anisotropic mechanisms (single ion and Dzyaloshinskii-Moriya) acting mainly in the Ru subsystem with RE ions playing a possible but not necessary role. It has been shown that a change of the details of the spin structure of the two sublattices ordering concomitantly at $T_{N1}$ is present between $T_{N1}$ and $T_{N2}$. This finding can be added to the formerly already proposed mechanisms of separate order of the two magnetic sublattices or of a change between 2D and 3D magnetic order. The present results should therefore be useful to develop realistic theoretical models to explain the presence and the mechanisms of two magnetic transitions in these double perovskites family. As seen on the example of the present study on $Sr_2YbRuO_6$, a prerogative for advancing further would demand better temperature dependent bulk measurements and neutron data on single crystals.

**Acknowledgement:** S. Sharma would like to acknowledge the Nanomission project of the Department of Science and Technology, India for providing postdoctoral fellowship during this work. DTA and ADH would like to acknowledge financial assistance from CMPC-STFC grant number CMPC-0910. The work of D.Kh. was funded by the Deutsche Forschungsgemeinschaft (DFG, German Research Foundation) - Project number 277146847 - CRC 1238. We would like to thank C.V. Tomy, R. Singh and Winfried Kockelmann for participating in the initial stage of this project on GEM and R. Singh for providing the sample. We thank F. Bernardini for his help during the WISH experiment and R. Stewart, G. Nilsen and Andrea Amorese for interesting discussions.

**Table I**: Selected bond angles (º) and bond lengths (Å) in paramagnetic (45 K) and AFM state (2 K) of $Sr_2YbRuO_6$.

| Bond angles (º) | 45 K | 2 K | Bond lengths (Å) | 45 K | 2 K |
|---|---|---|---|---|---|
| O1-Ru-O2 | 90.2(4) | 90.1(4) | Ru-O1 | 1.958(8) | 1.960(9) |
| O1-Ru-O3 | 90.9(4) | 90.8(4) | Ru-O2 | 1.959(9) | 1.957(1) |
| O2-Ru-O3 | 89.5(3) | 89.2(4) | Ru-O3 | 1.941(8) | 1.944(8) |
| O1-Yb-O2 | 87.9(4) | 91.9(4) | Yb-O1 | 2.164(8) | 2.163(9) |
| O1-Yb-O3 | 90.1(3) | 89.8(3) | Yb-O2 | 2.172(9) | 2.172(1) |
| O2-Yb-O3 | 89.1(3) | 89.3(3) | Yb-O3 | 2.182(8) | 2.180(8) |
| Ru-O1-Yb | 159.1(5) | 159.0(5) | Ru-Ru | 5.737(3) | 5.731(4) |
| Ru-O2-Yb | 157.7(5) | 158.1(5) | Ru-Yb | 4.054(1) | 4.045(1) |
| Ru-O3-Yb | 158.4(5) | 158.1(5) | Yb-Yb | 5.736(4) | 5.736(4) |

**Table II**
**(a) $Sr_2RERuO_6$:** The reported values of Weiss constant ($\theta_{CW}$), $T_N$ and the corresponding value of frustration index $f=|\theta_{CW}|/T_N$.

| Rare Earth (RE) | $\theta_{CW}$ (K) | $T_N$ (K) | $f=\theta_{CW}/T_N$ | References |
|---|---|---|---|---|
| Gd | -8.0 | 15.3 | 0.5 | [41] |
| Tb | -20 | 41 | 0.4 | [31] |
| Dy | -25 | 39.5, 36.5 | 0.7 | [8] |
| Ho | -20 | 20*, 36 | 0.6 | [42] |
| Er | -15.4 | 36 | 0.5 | [43] |
| Tm | -47 | 36 | 1.3 | [15] |
| Yb | -225 | 36, 42 | 5.35 | [15] |
| Lu | -353 | 30 | 11.7 | [6] |
|  |  | 27.2+, 29+ |  | [43] |
| Y | -292 | 25,32 | 9.1 | [44] |
| Y | −273.54 | 26,30 | 9.8 | [38] |
| Y | -336.6 | 24, 30 | 11.22 | [45] |

*For zero-field cool peak in the susceptibility, +Two transitions in the heat capacity

**(b) Ba$_2$RERuO$_6$:** The reported values of Weiss constant, T$_N$ and the corresponding value of frustration index.

| Rare Earth (RE) | θ$_{CW}$ | T$_N$ | f=θ$_{CW}$/T$_N$ | References |
|---|---|---|---|---|
| La | -383 | 29.5 | 13 | [46] |
| Pr | -133 | 117 | 1.14 | [47] |
| Nd | -35.5 | 27*, 58* | 0.61 | [32] |
| Ho | -19.9 | 22, 50 | 0.398 | [48] |
| Er | -14.6 | 40 | 0.365 | [49] |
| Tm | -34 | 42 | 0.81 | [15] |
| Yb | -181 | 48 | 3.78 | [15] |
| Lu | -630 | 35 | 18 | [6] |
| Y | -630 | 37 | 17 | [6] |
|   | -522 | 37$, 46$ | 16 | [50] |

* Two transitions in the magnetic susceptibility, but the heat capacity shows only one peak at 58 K.

$ Two transition in the magnetic susceptibility, but heat capacity shows only one peak at 36 K.

**Figure Captions:**

**Fig. 1:** (a) Rietveld refined XRD pattern of Sr$_2$YbRuO$_6$ at 300 K using the monoclinic space group $P112_1/n$. (b) Schematic representation of crystal structure at room temperature. The enlarged view of two dashed box regions is given in figure (c) and (d) respectively to clearly show the various bond lengths and bond angles in order to explain the possible magnetic interactions pathways. The local point symmetry of both Ru and Yb ion is triclinic (C$_i$) in the monoclinic crystal structure.

**Fig. 2:** (a) The *dc* magnetic susceptibility ($\chi_{dc}$) measured at various applied magnetic fields in zero-field cooled (ZFC) and field cooled (FC) conditions. The arrows indicate the magnetic transitions as $T_{N1}$ and $T_{N2}$ and the dashed line indicates the third weak anomaly $T^*$ near 10 K. The inset shows the enlarged view close to magnetic transitions. (b) Magnetization isotherms measured at various temperatures ranging from 5 to 300 K. The inset shows the enlarged view at lower fields data to show the hysteresis observed at 5 K and 30 K.

**Fig. 3** Real part of *ac* susceptibility ($\chi'$) measured with 10 Oe drive field in zero field cooled conditions at different frequencies ranging from 100 Hz to 10 kHz. The arrows indicate the magnetic transitions temperatures, $T_{N1}$ and $T_{N2}$. The insets (i) and (ii) represent the enlarged view near $T_{N1}$ and $T_{N2}$ respectively. The peak at $T_{N1}$ is frequency independent while the feature at $T_{N2}$ is slightly frequency dependent. Refer to text for details.

**Fig. 4:** Heat capacity as a function of temperature in zero and 2 Tesla applied magnetic field.

**Fig. 5:** Zero-field μSR spectra measured at various temperatures. The experimental data are shown by the symbols and the solid red line shows fit to the data using an exponential decay function.

**Fig. 6:** The temperature dependent parameters obtained from the fit to μSR spectra as a function of temperature. The initial muon asymmetry ($A_0$) and relaxation rate ($\lambda$) are plotted on right and left *y*-scale with linked *x*-scale.

**Fig. 7:** Thermal evolution of magnetic reflections below $T_{N1}$. The plotted temperature range is 2 K to 45 K. The black arrows indicate the onset of the magnetic Bragg reflections $T_{N1}$ and the red arrows highlight the changes in the diffracted intensity at $T_{N2}$. A small but clear enhancement in diffracted intensity below $T_{N2}$ (red arrow) is evident in the graph.

**Fig. 8:** (a-d) The temperature variation of the integrated intensity of various magnetic reflections extracted from the difference curve. Error bars are smaller than the symbol size. Two components are clearly visible as shown by the two red lines and highlighted by the shaded regions. The second component starts growing below $T_{N2}$. (Refer to the text for details).

**Fig. 9:** Rietveld refined NPD patterns collected at (a) 100 and (b) 2 K. Two series of tick marks in (b) correspond to the nuclear (upper-green) and magnetic (lower-red) Bragg reflections. The observed, calculated intensities and difference are plotted as solid circles, solid line and bottom line, respectively. The inset in (b) shows the fitted difference data (2 K–45 K) using just the magnetic model.

**Fig. 10:** (a) The magnetic structure of $Sr_2YbRuO_6$ for $k$ = (0 0 0). The $Yb^{3+}$ and $Ru^{5+}$ moments are shown by cyan (small) and red colours (large) arrows, respectively. (b) The spherical coordinate setting used in the present work.

**Fig. 11:** Thermal variation of (a) $Ru^{5+}$ moments, (b) $Yb^{3+}$ moments, (c) moments angle $R_\phi$ (with respect to $x$-axis/$a$-axis) while $R_\theta$ = 90° and (d) the Normalized moments of $Yb^{3+}$ and $Ru^{5+}$. The vertical black dashed line corresponds to $T_{N2}$.

**Fig. 12:** The schematic orbital diagrams of $Ru^{5+}$, $O^{2-}$ and $Yb^{3+}$, (a) Mechanism of AFM Ru-Ru exchange. Grey are Ru $t_{2g}$ orbitals, blue are Oxygen 2p orbitals and (b) Schematic illustration of FM Ru-Yb exchange interaction. Grey is Ru $t_{2g}$ orbital and purple Yb $\Gamma_6$ (CEF ground state) orbital taken from ref. [34].

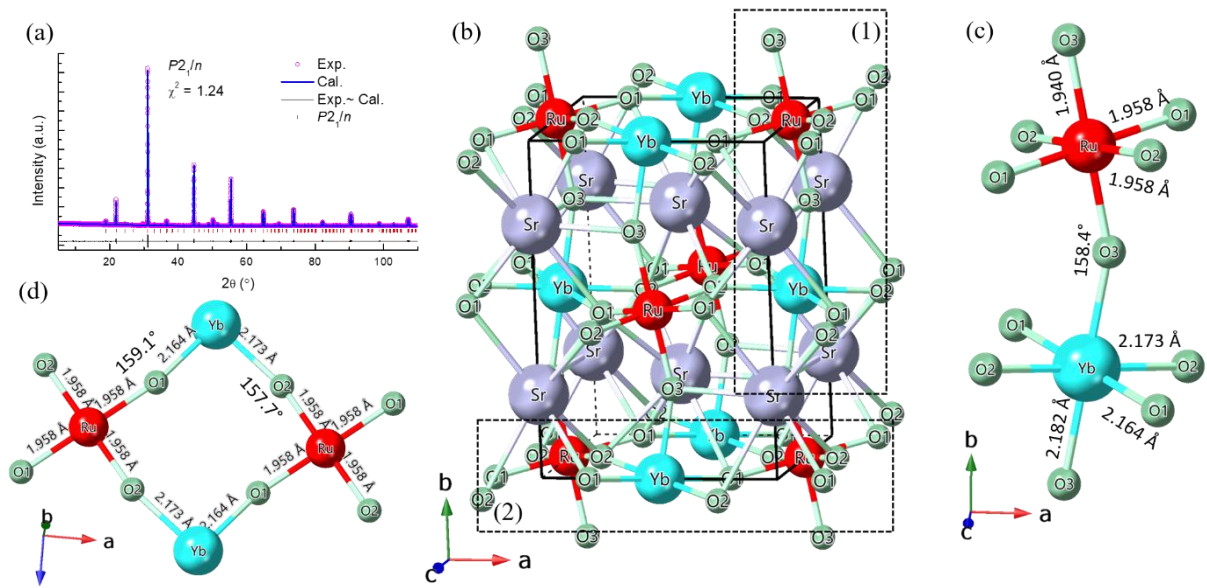

**Fig. 1:** (a) Rietveld refined XRD pattern of $Sr_2YbRuO_6$ at 300 K using the monoclinic space group $P112_1/n$. (b) Schematic representation of crystal structure at room temperature. The enlarged view of two dashed box regions is given in figure (c) and (d) respectively to clearly show the various bond lengths and bond angles in order to explain the possible magnetic interactions pathways. The local point symmetry of both Ru and Yb ion is triclinic ($C_i$) in the monoclinic crystal structure.

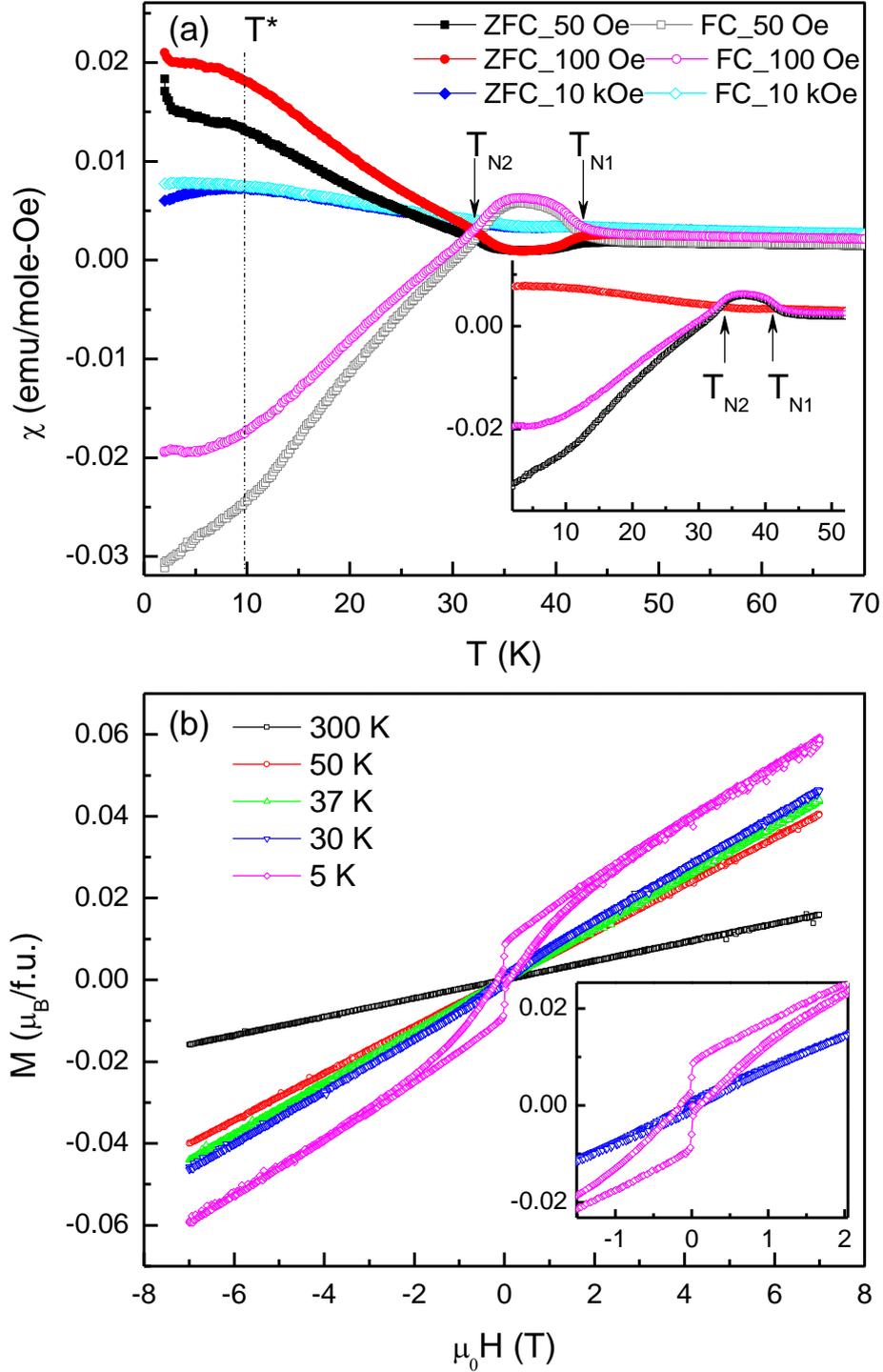

**Fig. 2:** (a) The *dc* magnetic susceptibility ($\chi_{dc}$) measured at various applied magnetic field in zero-field cooled (ZFC) and field cooled (FC) conditions. The arrows indicate the magnetic transitions as $T_{N1}$ and $T_{N2}$ and the dashed line indicates the third weak anomaly $T^*$ near 10 K. The inset shows the enlarge view close to magnetic transitions. (b) Magnetization isotherms measured at various temperatures ranging from 5 to 300 K. The inset shows the enlarge view at lower fields to show the hysteresis observed at 5 K and 30 K.

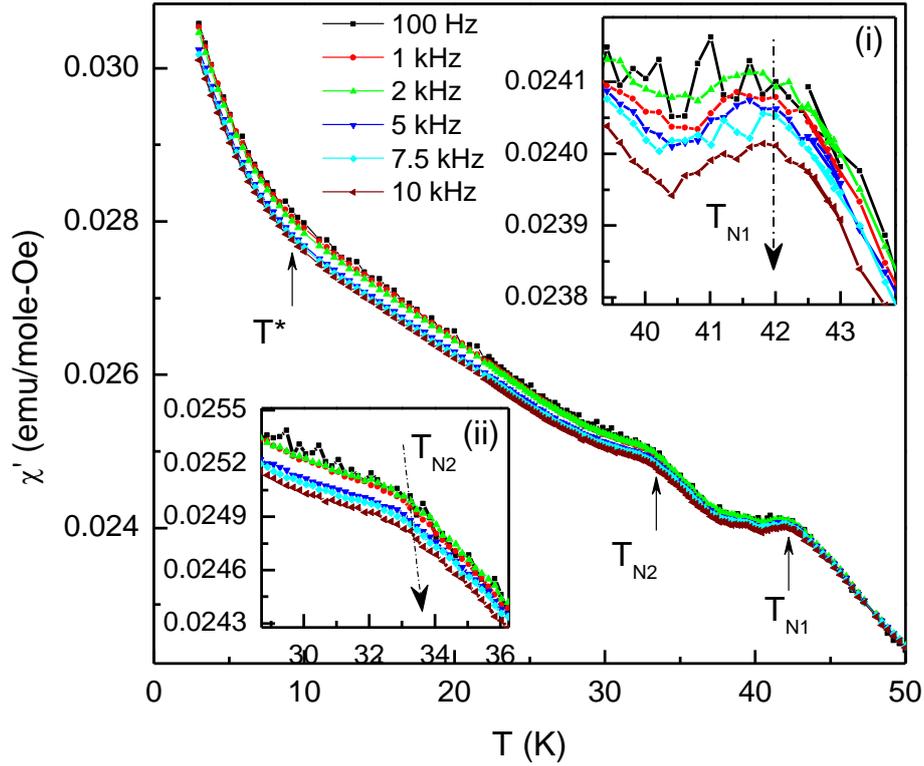

**Fig. 3** Real part of *ac* susceptibility ($\chi'$) measured with 10 Oe drive field in zero field cooled conditions at different frequencies ranging from 100 Hz to 10 kHz. The arrows indicate the magnetic transitions temperatures, $T_{N1}$ and $T_{N2}$. The insets (i) and (ii) represent the enlarged view near $T_{N1}$ and $T_{N2}$ respectively. The peak at $T_{N1}$ is frequency independent while the feature at $T_{N2}$ is slightly frequency dependent. Refer to text for details.

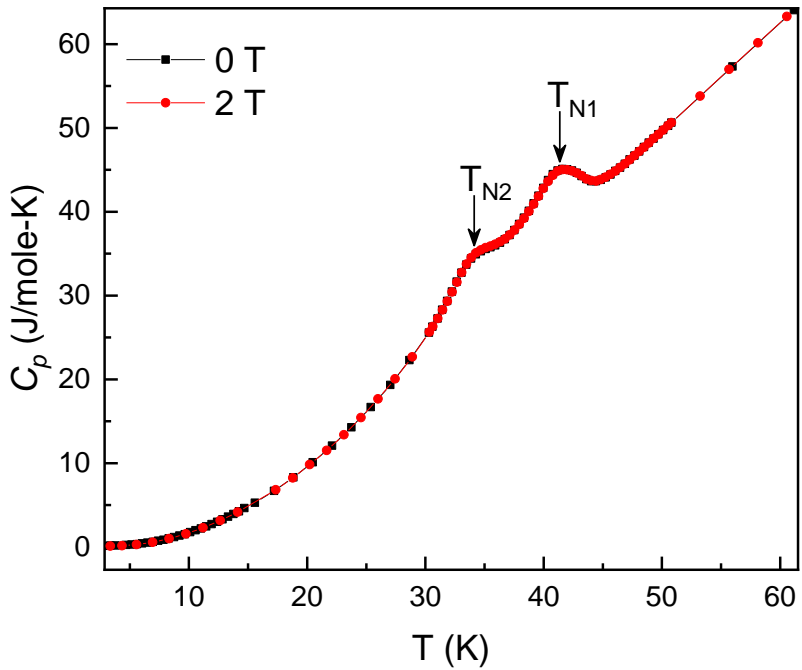

**Fig. 4:** Heat capacity as a function of temperature in the presence of 0 and 2 Tesla applied magnetic field.

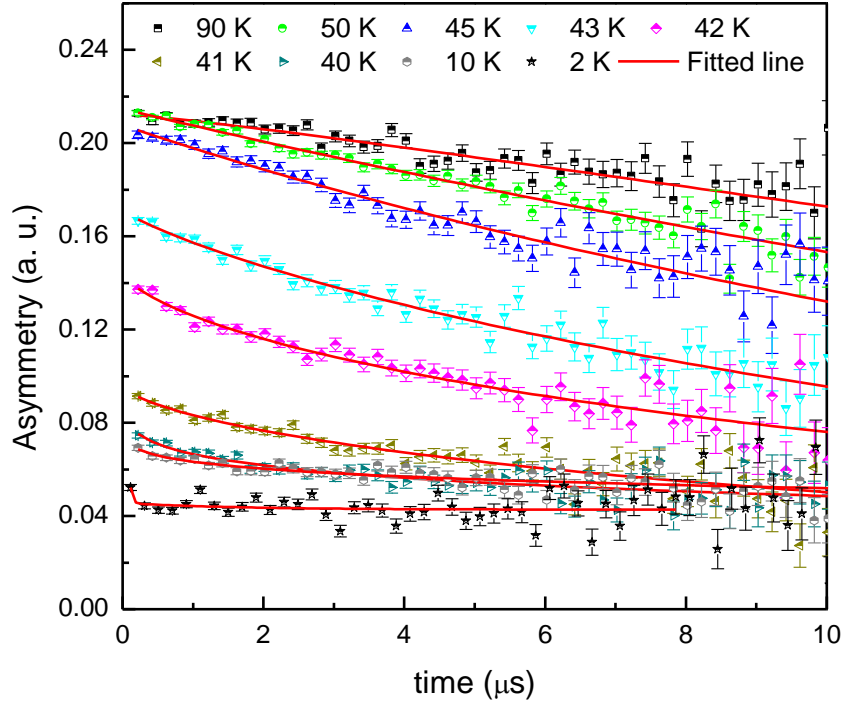

**Fig. 5:** Zero-field μSR spectra measured at various temperatures. The experimental data are shown by the symbols and the solid red lines show the fit to the data using an exponential decay function.

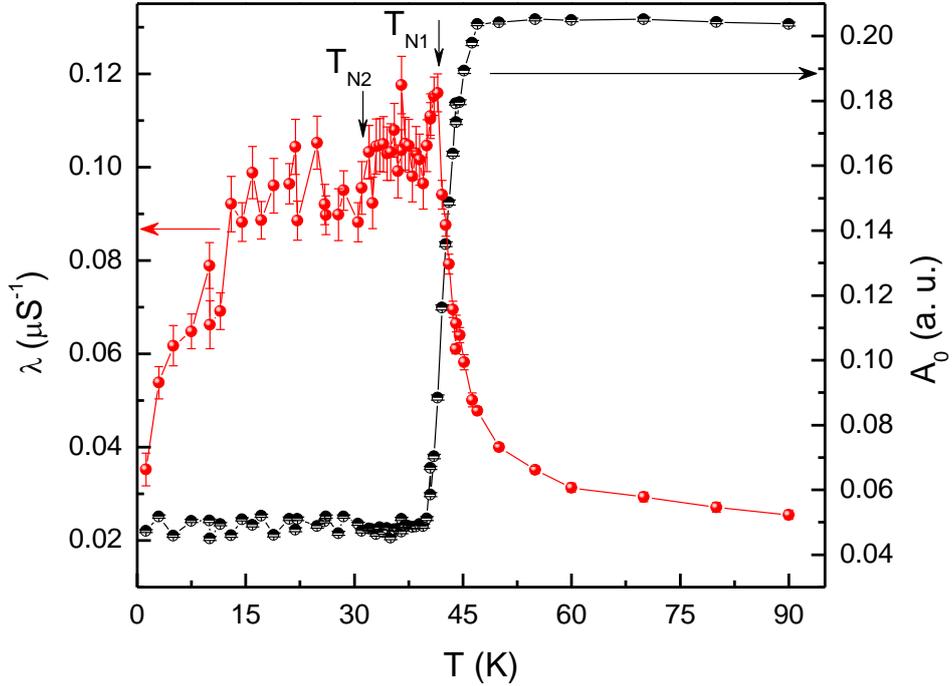

**Fig. 6:** The temperature dependent parameters obtained from the fit to μSR spectra as a function of temperature. The initial muon asymmetry ($A_0$) and relaxation rate ($\lambda$) are plotted on right and left *y*-scale with linked *x*-scale.

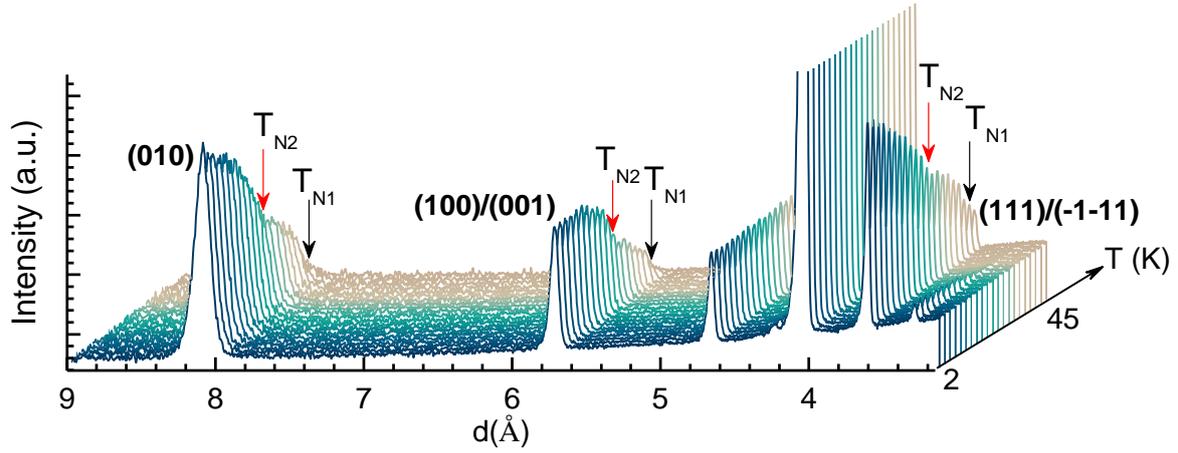

**Fig. 7:** Thermal evolution of magnetic reflections below $T_{N1}$. The plotted temperature range is 2 K to 45 K. The black arrows indicate the onset of the magnetic Bragg reflections $T_{N1}$ and the red arrows highlight the changes in the diffracted intensity at $T_{N2}$. A small but clear enhancement in diffracted intensity below $T_{N2}$ (red arrow) is evident in the graph.

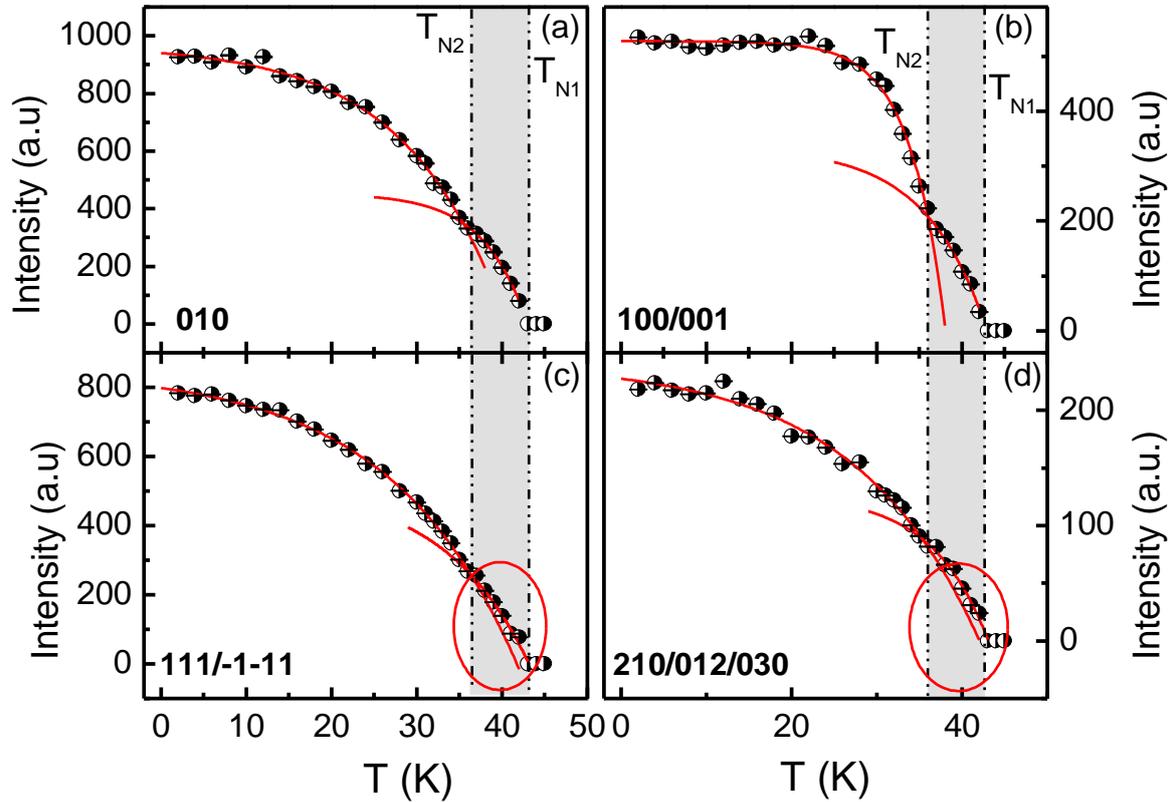

**Fig. 8:** (a-d) The temperature variation of the integrated intensity of various magnetic reflections extracted from the difference curve. Error bars are smaller than the symbol size. Two components are clearly visible as shown by the two red lines and highlighted by the shaded regions. The second component starts growing below $T_{N2}$. (Refer to the text for details).

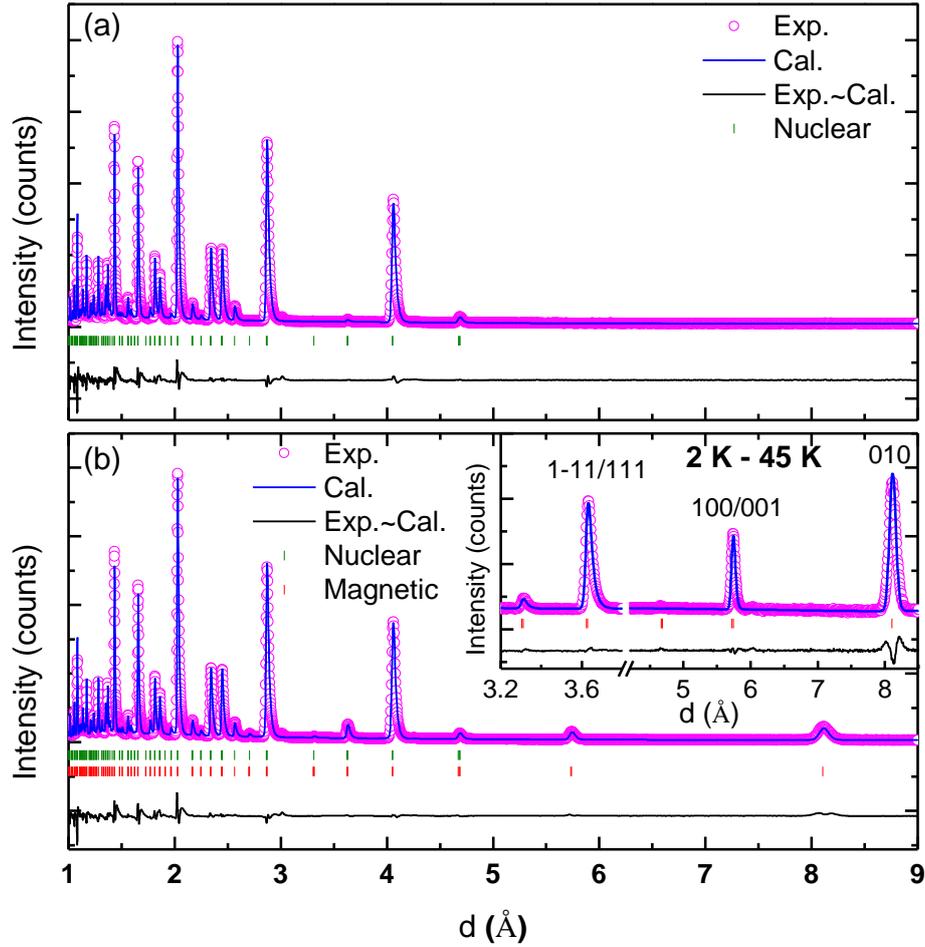

**Fig. 9:** Rietveld refined NPD patterns collected at (a) 100 and (b) 2 K. Two series of tick marks in (b) correspond to the nuclear (upper-green) and magnetic (lower-red) Bragg reflections. The observed, calculated intensities and difference are plotted as solid circles, solid line and bottom line, respectively. The inset in (b) shows the fitted difference data (2 K–45 K) using just the magnetic model.

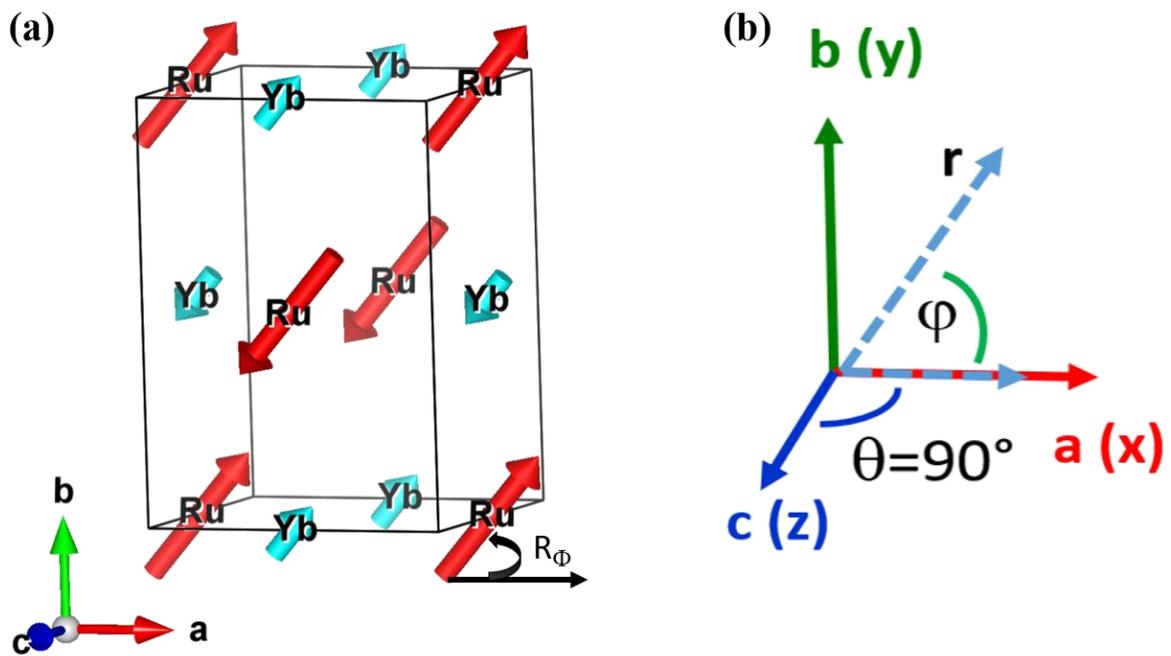

**Fig. 10:** (a) The magnetic structure of $Sr_2YbRuO_6$ for $k = (0\ 0\ 0)$. The $Yb^{3+}$ and $Ru^{5+}$ moments are shown by cyan (small) and red colours (large) arrows, respectively. (b) The spherical coordinate setting used in the present work.

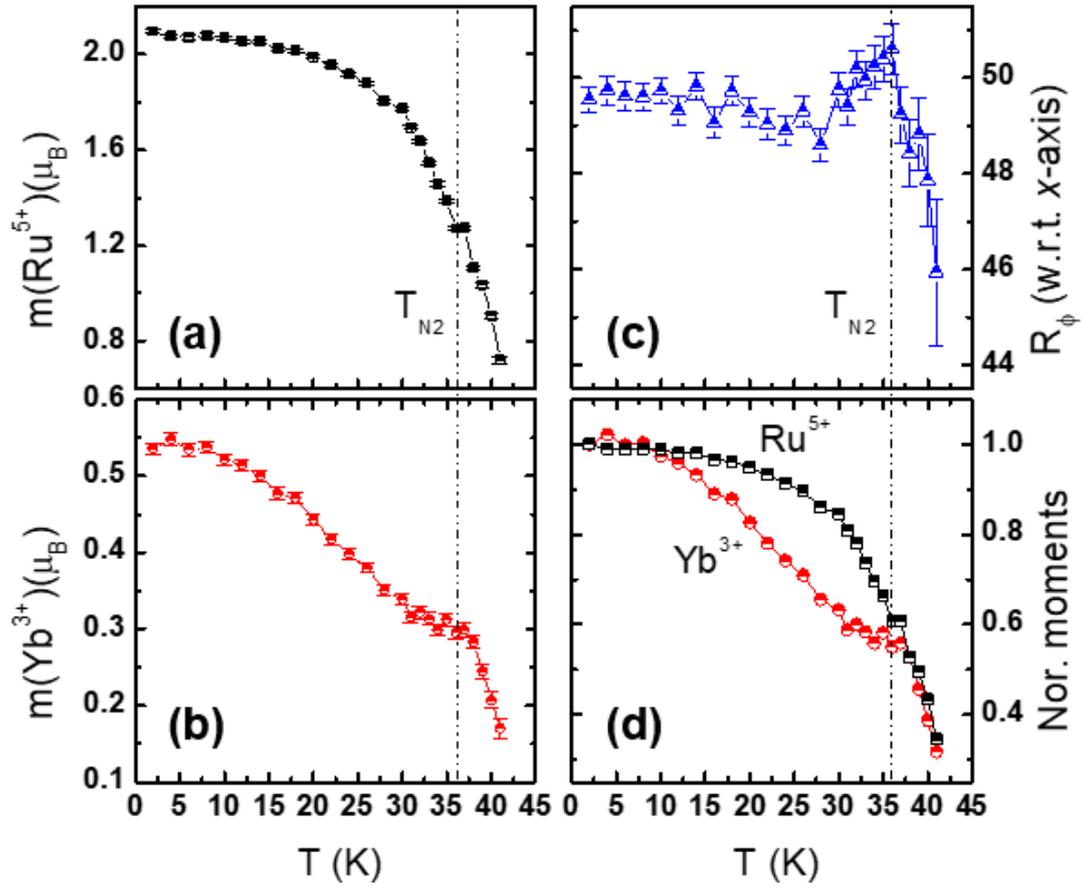

**Fig. 11:** Thermal variation of (a) $Ru^{5+}$ moments, (b) $Yb^{3+}$ moments, (c) moments angle $R_\phi$ (with respect to x-axis/a-axis) while $R_\theta = 90°$ and (d) the Normalized moments of $Yb^{3+}$ and $Ru^{5+}$. The vertical black dashed line corresponds to $T_{N2}$.

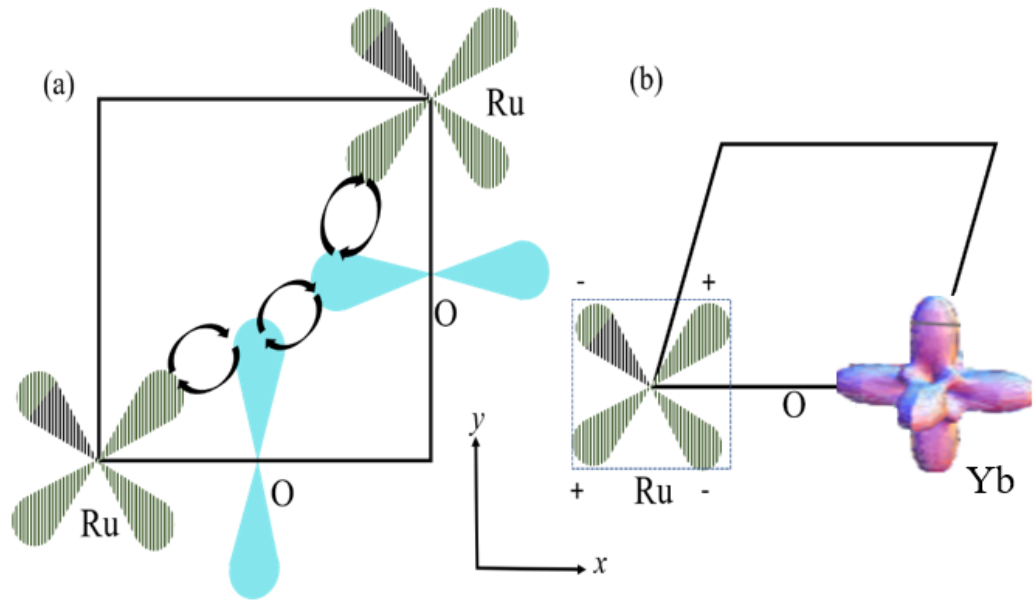

**Fig. 12:** The schematic orbital diagrams of $Ru^{5+}$, $O^{2-}$ and $Yb^{3+}$, (a) Mechanism of AFM Ru-Ru exchange. Grey are Ru $t_{2g}$ orbitals, blue are Oxygen 2p orbitals and (b) Schematic illustration of FM Ru-Yb exchange interaction. Grey is Ru $t_{2g}$ orbital and purple Yb $\Gamma_6$ (CEF ground state) orbital taken from ref. [34].

# Supplementary Materials

**Investigation of the magnetic ground state of the ordered double perovskite $Sr_2YbRuO_6$: a two transitions**

Shivani Sharma[1,2], D. T. Adroja[1,3], C. Ritter[4], D. Khalyavin[1], P. Manuel[1], Gavin B. G. Stenning[1],

A. Sundaresan[2], A. D. Hillier[1], P.P. Deen[5]. D. I. Khomskii[6], S. Langridge[1],

[1]*ISIS facility, Rutherford Appleton Laboratory, Chilton Oxon, OX11 0QX*
[2]*Jawaharlal Nehru Centre for Advanced Scientific Research, Jakkur, Bangalore, India*
[3]*Highly Correlated Matter Research Group, Physics Department, University of Johannesburg, Auckland Park 2006, South Africa*
[4]*Institut Laue-Langevin, 71 Avenue des Martyrs, CS 20156, 38042, Grenoble Cedex 9, France*
5*European Spallation Source ERIC, 22363 Lund, Sweden*
*Niels Bohr Institute, University of Copenhagen, Denmark*
[6]*II Physikalisches Institut, Universitaet zu Koeln, Zuelpicher Str. 77, 50937 Koeln, Germany*
(Date: 18-09-2020)

Fig.1S shows the magnetic structure of $Sr_2YbRuO_6$ plotted for comparison in the two different settings found in the literature.

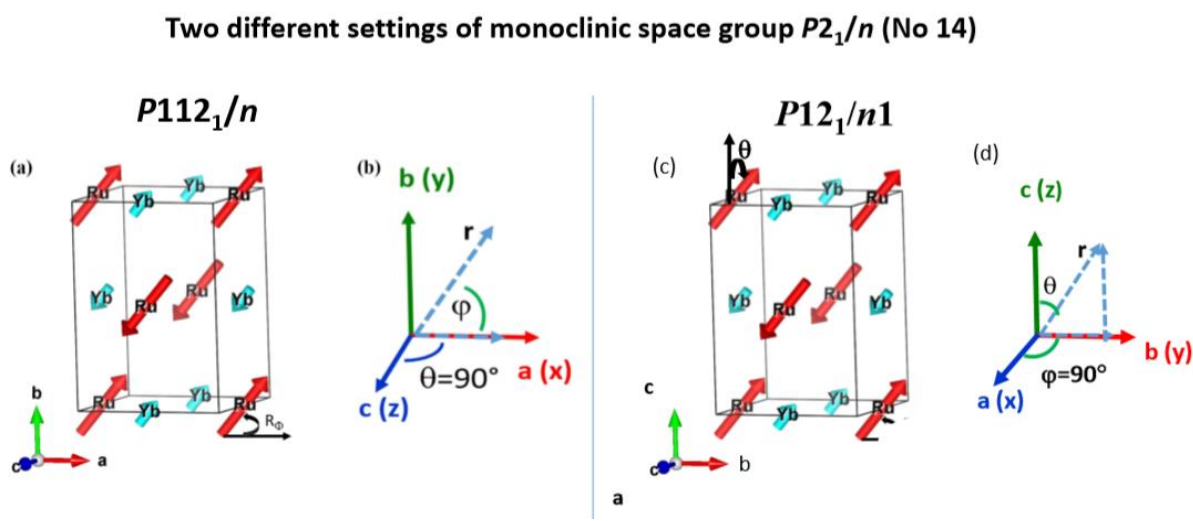

Fig.1S Magnetic structure of $Sr_2YbRuO_6$ plotted in two different settings (a, b) $P112_1/n$ setting used in the present paper and (c, d) $P12_1/n1$ used by Doi et al, ref. [1]

**References:**

[1] Y. Doi, Y. Hinatsu, A. Nakamura, Y. Ishii, and Y. Morii, J. Mater. Chem. **13**, 1758 (2003).